\documentclass{article}

\usepackage{arxiv}

\usepackage[utf8]{inputenc} 
\usepackage[T1]{fontenc}    
\usepackage{hyperref}       
\usepackage{url}            
\usepackage{booktabs}       
\usepackage{amsfonts}       
\usepackage{nicefrac}       
\usepackage{microtype}      
\usepackage{lipsum}
\usepackage{graphicx}
\usepackage{enumerate} 
\usepackage{paralist}
\usepackage{listings}
\usepackage{multirow}
\usepackage{subfig}
\usepackage{graphicx}
\usepackage{tikz}
\usepackage{url}
\usepackage{xcolor}
\usepackage{soul}
\usepackage{xargs}
\usepackage{balance}
\usepackage{multirow}
\usepackage{enumitem}
\usepackage{footmisc}
\usepackage{parcolumns}
\usepackage{float}
\graphicspath{ {./images/} }

\usepackage{listings}
\usepackage{graphicx}
\usepackage{amsmath}
\usepackage{amssymb}
\usepackage{graphicx}
\usepackage{color,soul}
\usepackage{url}
\newcommand{\orcid}[1]{\href{https://orcid.org/#1}{\includegraphics[height=10pt]{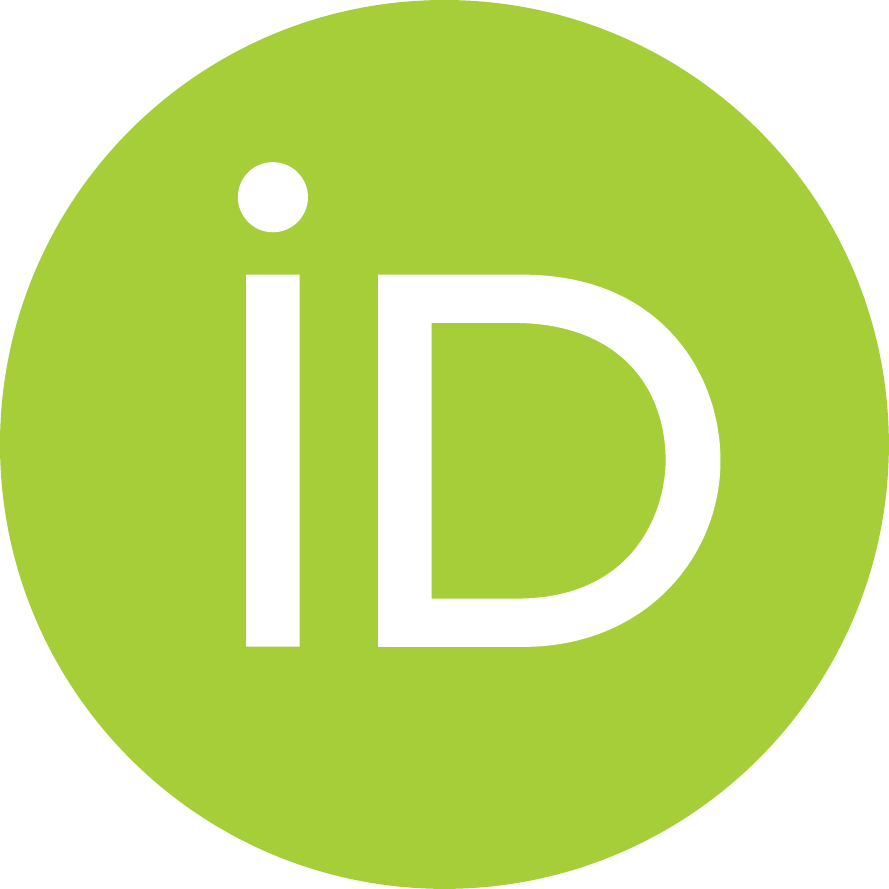}}}

\title{A Case Study of LLVM-Based Analysis for Optimizing SIMD Code Generation}

\author{
Joseph Huber \\
    Oak Ridge National Laboratory \\
    \texttt{huberjn@ornl.gov}
\And
Weile Wei\orcid{0000-0002-3065-4959} \\
  Louisiana State University\\
  \texttt{wwei9@lsu.edu} \\
\And
Giorgis Georgakoudis \\
    Lawrence Livermore National Laboratory \\
    \texttt{georgakoudis1@llnl.gov}
\And
Johannes Doerfert \\
  Argonne National Laboratory\\
  \texttt{jdoerfert@anl.gov} \\
\And
Oscar Hernandez\orcid{0000-0002-5380-6951} \\
  Oak Ridge National Laboratory\\
  \texttt{oscar@ornl.gov} \\
}

\begin{document}
\maketitle
\begin{abstract}
This paper presents a methodology for using LLVM-based tools to tune the DCA++ (dynamical cluster approximation) application that targets the new ARM A64FX processor. The goal is to describe the changes required for the new architecture and generate efficient single instruction/multiple data (SIMD) instructions that target the new Scalable Vector Extension instruction set.
During manual tuning, the authors used the LLVM tools to improve code parallelization by using OpenMP SIMD, refactored the code and applied transformation that enabled SIMD optimizations, and ensured that the correct libraries were used to achieve optimal performance. By applying these code changes, code speed was increased by 1.98$\times$ and 78 GFlops were achieved on the A64FX processor. The authors aim to automatize parts of the efforts in the OpenMP Advisor tool, which is built on top of existing and newly introduced LLVM tooling.
\end{abstract}


\footnotetext{This manuscript has been co-authored by UT-Battelle, LLC under Contract No. DE-AC05-00OR22725 with the U.S. Department of Energy.  The United States Government retains and the publisher, by accepting the article for publication, acknowledges that the United States Government retains a non-exclusive, paid-up, irrevocable, world-wide license to publish or reproduce the published form of this manuscript, or allow others to do so, for United States Government purposes. The Department of Energy will provide public access to these results of federally sponsored research in accordance with the DOE Public Access Plan (http://energy.gov/downloads/doe-public-access-plan).}

\section{Introduction}

Program analysis tools are important in 
helping users understand, improve, and port their applications to new platforms. This is crucial for applications that need tuning and 
significant code restructuring to exploit new types of hardware devices, such as single instruction/multiple data (SIMD) units and accelerators. Compiler-based tools are crucially important for identifying opportunities to improve application codes as the compiler generates code for different architectures.
In particular, the LLVM compiler is an open-source compiler that provides a set of tools for the static analysis and feedback of application code. Static program analysis information can be combined with dynamic information (profile-based) to filter the large amount of information produced by the compiler so that users can focus on the most frequently executed regions of their code.

This paper presents a methodology for using LLVM-based tools to tune an application to generate efficient SIMD instructions that target the new ARM A64FX processor, as well as describes what is required to achieve good performance.

\section{Case Study: Porting DCA++ to Wombat}
This section describes the authors' experiences in porting the DCA++ (dynamical cluster approximation) application to the Wombat\footnote[1]{Wombat: \url{www.olcf.ornl.gov/olcf-resources/compute-systems/wombat/}} cluster, an ARM-based heterogeneous cluster at Oak Ridge National Laboratory. This section presents a methodology for using LLVM-based tools to tune the DCA++ application targeting the ARM A64FX and ThunderX2 processors. The goal is to describe what changes are required for the new architecture and generate efficient SIMD instructions that target the new Scalable Vector Extension (SVE) instruction set available in the A64FX processors based on LLVM-based tools information.

\subsection{Evaluation Environment} 
The case study used the Wombat test bed with 24 compute nodes.
Sixteen compute nodes are based on the Fujitsu A64FX processor with SVE and a theoretical peak performance of 3.3792 TFlops.
Each A64FX node has one processor socket with 32 GB of second-generation High-Bandwidth Memory (HBM2).
The A64FX-equipped nodes do not have additional Double Data Rate (DDR) memory.
Eight compute nodes have two ThunderX2 processors with NEON vector instructions and a theoretical peak performance of 560 GFlops.
The ThunderX2 nodes have 256 GB of DDR4 RAM and a 480 GB solid-state drive for node-local storage.
All nodes are connected with Enhanced Data Rate InfiniBand (100 Gbit/s).
The compilers on the system are the ARM 20.3 compilers and the Clang upstream compiler, which is based on Clang 12. 
The scientific libraries available on Wombat are the ARM Performance Libraries (APL) version 20.3.


\subsection{DCA++} 
Quantum Monte Carlo (QMC) solver applications are popular tools essential to the US Department of Energy-supported scientific software.
This paper studies one cutting-edge QMC application called the DCA++ algorithm. DCA++~\cite{hahner2020dca++} implements quantum cluster
algorithms to solve quantum many-body problems in condensed matter physics.
DCA++ is a highly scalable and performant scientific software written in
modern C++ and has been ported to various high-performance computing architectures, including IBM Power9, x86\_64, ThunderX2, and ARM A64FX~\cite{dca_hpx_2020}. 
The DCA++ software currently integrates three different programming
models---message passing interface (MPI), Compute Unified Device Architecture (CUDA), and  High Performance ParalleX (HPX)/C++ threading---together with numerical libraries (e.g., Basic Linear Algebra Subprograms [BLAS], Linear Algebra Package [LAPACK], and MAGMA) to expose the parallel computation structure.

Wei et al.~\cite{dca_hpx_2020} reported that DCA++ with the HPX run time 
system~\cite{Kaiser2020} has produced a 20\% run time speedup over the one with C++ 
standard threading support. 
The speedup is primarily due to the faster thread context
switching and reduced scheduler synchronization overheads in the HPX run time.
Moreover, Autonomic Performance Environment for Exascale (APEX)~\cite{huck2015autonomic} is an in situ profiling and adaptive tuning framework to the HPX run time system that can capture operating system and hardware
system performance data through various interfaces, such as Performance Application Programming Interface (PAPI)~\cite{papi2010}.
Because APEX is highly integrated into the HPX run time, for HPX-supported applications, users can
easily capture PAPI counter information (e.g., level 2 data cache misses, vector/SIMD instructions, floating point instructions) through HPX function annotation.
The overhead introduced by APEX profiling is as 
low as $\sim$1\%~\cite{diehl2021performance}
compared with the overall application run time.

In DCA++, the QMC solver is the most computation-intensive
unit that models strongly
correlated electron systems~\cite{dca_hpx_2020}.
Computation on the QMC solver is parallelized by using a multithreading scheme that comprises \texttt{walker} (i.e., producer) and \texttt{accumulator} (i.e., consumer) tasks. Each task runs on an independent thread. There are multiple \texttt{walker}s running concurrently.
Each \texttt{walker} is responsible for a Monte Carlo (MC) update (sampling from the Markov
chain), and then an \texttt{accumulator} is popped from the head of the \texttt{accumulator waiting queue} to
compute an MC measurement from the \texttt{walker}. When each \texttt{accumulator} finishes
its accumulation measurement, it is pushed back to the end of the queue. The walker-accumulator synchronization is
managed by the synchronization primitives \texttt{mutex} and
\texttt{conditional\_variable}. 

\subsection{Baseline Performance}
The following experiments
compare DCA++'s performance on Wombat by using its A64FX and ThunderX2 nodes.
The performance is measured using 48 \texttt{accumulator}s and 48 \texttt{walkers} and using 100,000 measurements, which is a representative scientific simulation case in production.
On A64FX, DCA++ is built with two different configuration
settings: SVE vectorization and SVE-disabled.
The SVE vectorization version of DCA++ means that 
DCA++ is built with SVE compiler flags enabled and 
vectorized loops, and it uses the APL optimized for SVE (i.e., LAPACK, BLAS, Fastest Fourier Transform in the West [FFTW]). 
The SVE compiler flags are set to
``-DNDEBUG -fsimdmath -fopenmp -O3 -mcpu=a64fx''
The SVE-disabled version means that DCA++ is built 
with original DCA++ code and open-source scientific
libraries, including Netlib-LAPACK and FFTW.
Similarly, on ThunderX2, DCA++ is built with two 
different configurations: with NEON and NEON disabled.

\begin{figure}
    \centering
	\includegraphics[width=0.75\columnwidth, trim=0cm 0cm 0cm 0cm, clip]{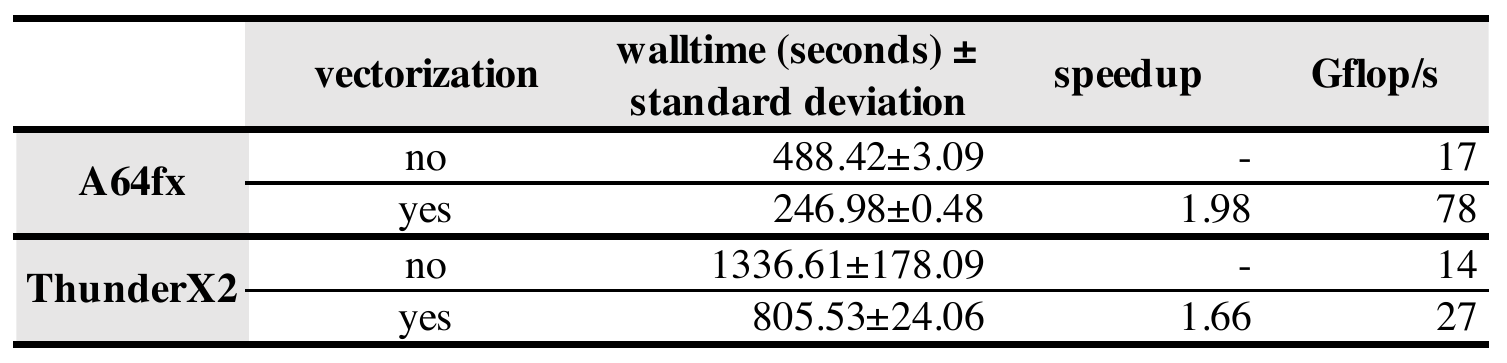}
	\caption{DCA++ execution time.}
	\label{fig:timing}
\end{figure}

Figure~\ref{fig:timing} shows DCA++ execution time
on A64FX and ThunderX2 architectures.
On A64FX, the SVE vectorization version of DCA++ performs $\sim$2$\times$ faster than the 
SVE-disabled version. On ThunderX2, the NEON version of DCA++ is observed to be $\sim$1.66$\times$ faster than the NEON-disabled version. 
Noticeably, the SVE vectorization version of DCA++ on A64FX has $\sim$3.3$\times$ speedup over
the NEON version on ThunderX2.
Meanwhile, the NEON version on ThunderX2 is measured to
have $\sim$27 GFlops, and the SVE vectorization version of DCA++ on A64FX reached $\sim$78 GFlops ($\sim$2.8$\times$).

Thes results show the performance gains of DCA++ due to the peak performance improvements of the A64FX processor (e.g., 500 GFlops for ThunderX2 vs. 2.5 TFlops for A64FX).

\begin{figure}[ht]
    \centering
	\includegraphics[width=0.9\columnwidth, trim=0.1cm 0.1cm 0.1cm 1cm, clip]{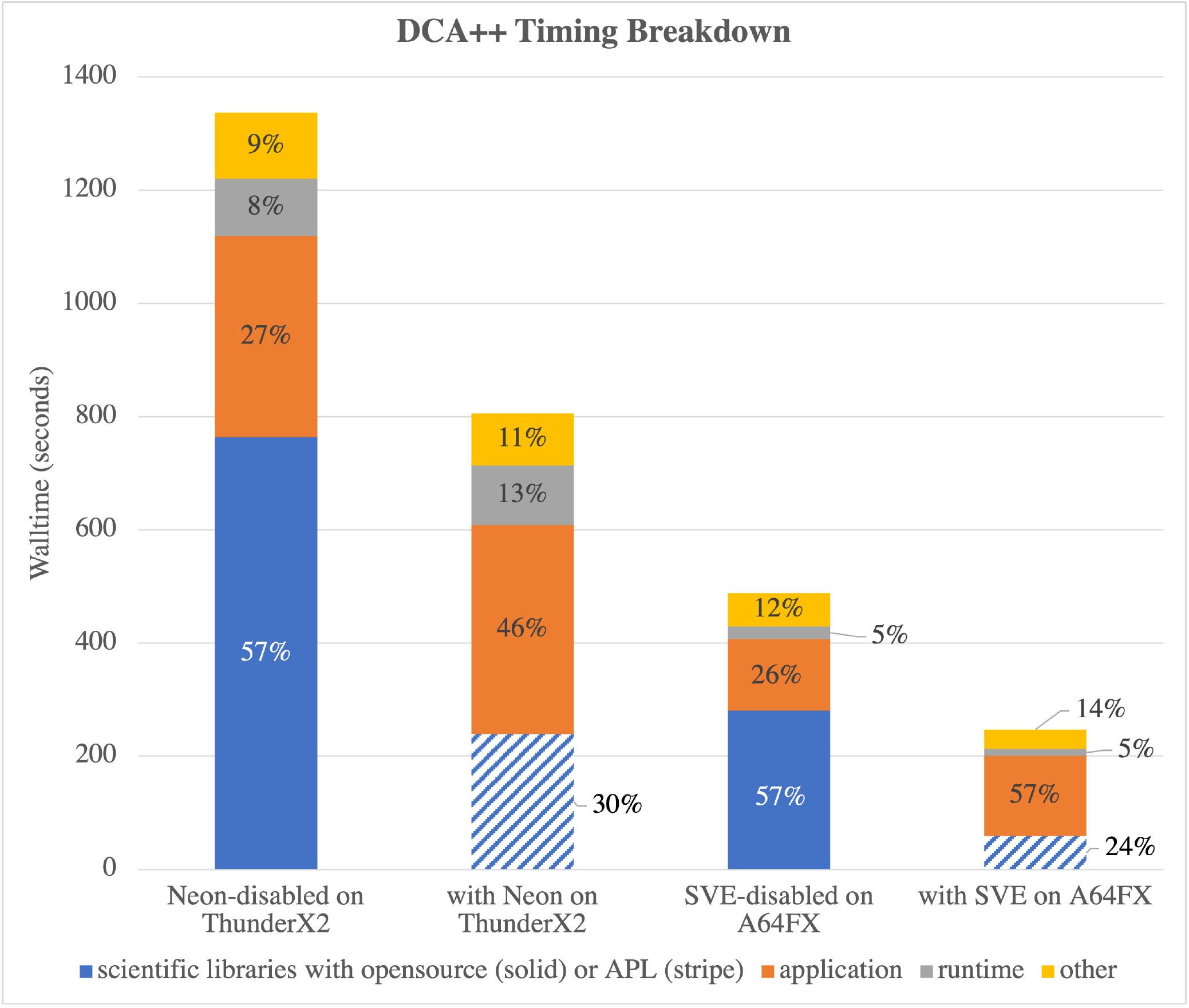}
	\caption{DCA++ timing breakdown.}
	\label{fig:timing_breakdown}
\end{figure}

Figure~\ref{fig:timing_breakdown} shows the breakdown of DCA++ execution time into four categories: application, scientific libraries, 
HPX run time, and other activities. 
Each category only considers functions that have more than 1\% overhead shown in the final 
profiling report generated from \texttt{perf}, a Linux built-in performance profiling tool.
The application category includes custom modules
developed in the DCA++ source code. 
The HPX run time category represents necessary scheduling and coordination efforts in HPX threads manager. The scientific libraries category captures routines from external numerical libraries, 
such as BLAS, LAPACK, FFTW, and math routines. The other activities category summarizes all other functions that have less than 1\% overhead in the
final profiling report.

Several observations were made from the timing breakdown shown in Fig.~\ref{fig:timing_breakdown}.
\begin{enumerate}
    \item With SVE vectorization or NEON optimization,
    the dominant percentage of the overall execution
    time is shifted from the external scientific libraries to the application source code.
    For example, on A64FX, the percentage of application time in the SVE-disabled vectorization version of DCA++ is 26\%, whereas the percentage of application time in the SVE version is 57\%.
    A similar percentage shift
    is also observed on ThunderX2 comparisons. 
    In other words, with APL (SVE vectorization on A64FX or NEON optimization on ThunderX2), less time is spent on scientific libraries because
    APL are particularly optimized 
    on targeting platforms.
    
    \item The HPX run time library imposes minimal overhead
    to the overall program execution.
    The overhead is primarily due to a lack of sufficient parallelism from the application so that
    some HPX worker threads in the kernel level 
    are spinning and waiting for user-level tasks. 
\end{enumerate}

\begin{figure}
    \centering
	\includegraphics[width=0.9\columnwidth, trim=0cm 0cm 0cm 0cm, clip]{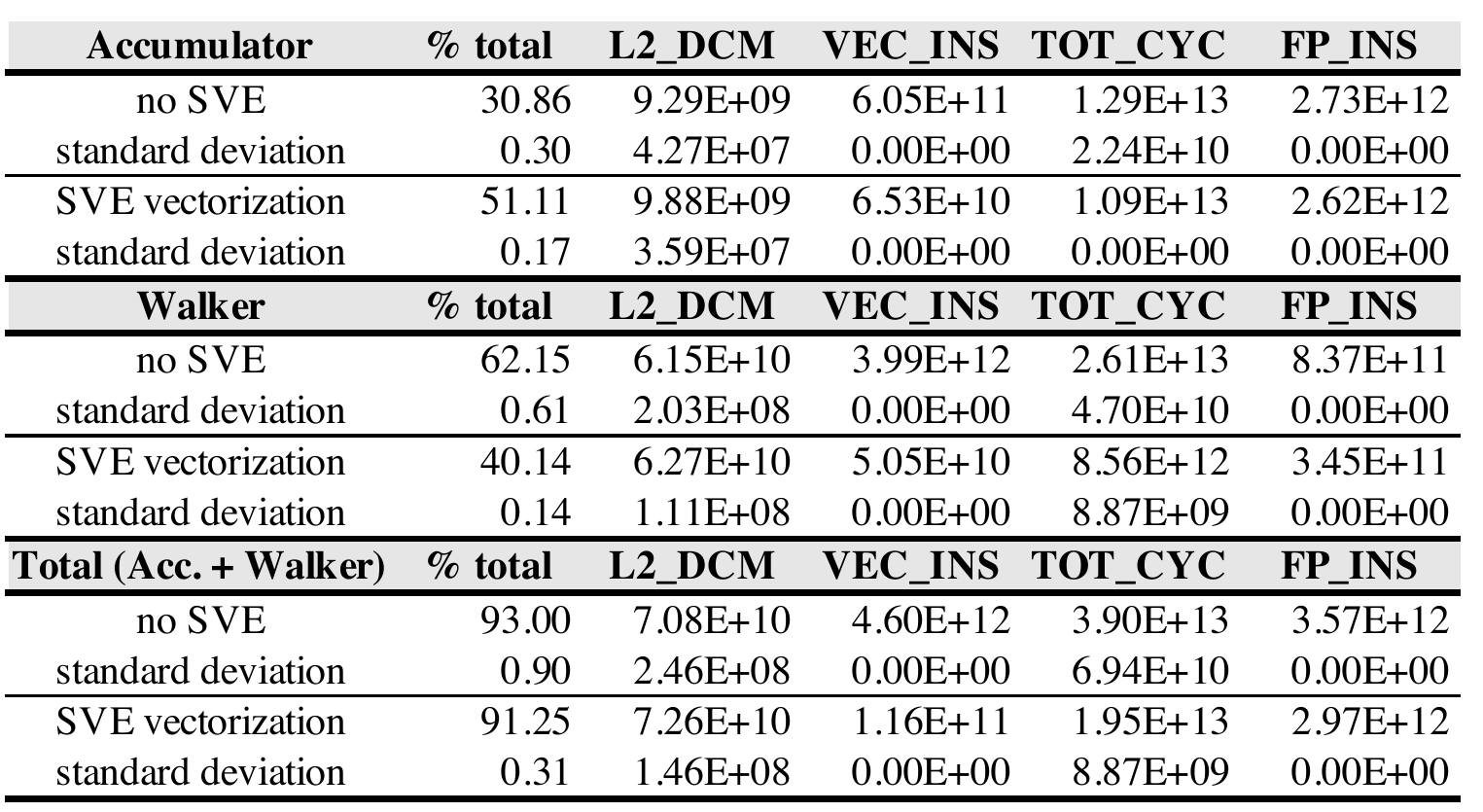}
	\caption{PAPI counter for DCA++ runs on A64FX.}
	\label{fig:papi_counter_A64FX}
\end{figure}

Further investigation using hardware performance counters is shown in Fig.~\ref{fig:papi_counter_A64FX}. Here, \texttt{hpx::annotated\_function()} is used to wrap \texttt{accumulator} 
and \texttt{walker} tasks so that their activities (i.e., 
timing information and PAPI counters) can be distinguished in the final profiling report generated from the HPX-APEX profiling tool. 
Figure ~\ref{fig:papi_counter_A64FX} shows that 
the total execution time of \texttt{accumulator} 
and \texttt{walker} takes the majority of the overall
program execution time ($\sim$93.00\% in the SVE-disabled version 
and $\sim$91.25\% in SVE vectorization version).
Several observations were made from Fig.~\ref{fig:papi_counter_A64FX}.

\begin{enumerate}
    \item The SVE-disabled version of DCA++ on A64FX has nearly $\sim$40$\times$ higher VEC\_INC, 2$\times$ higher TOT\_CYC, and 1.2$\times$
    higher FP\_INS than 
    the SVE vectorization version, where
    VEC\_INC is vector/SIMD instructions, TOT\_CYC is total cycles,
    and FP\_INC is floating point instructions. The authors noticed that by using the optimized libraries, the application uses less vector and floating point SVE instructions. Because SVE has
    wider 512 bit width, fewer 
    vector instructions are needed in the computation than NEON, which has 128 bit width. Also, the SVE has a more powerful instruction set that uses
    fewer instructions for the same operation.
    \item The L2\_DCM (L2 data cache misses) does not change with the SVE optimized version because the SVE optimization does not impact overall memory access patterns. Access to HBM2 remained constant in both versions.
    \item Using SVE vectorization on DCA++ shifts timing percentages between \texttt{accumulator} and \texttt{walker} in overall program execution. 
   To perform efficient matrix-related operations, the implementation of \texttt{walker} extensively uses DGEMM routines, which are provided by the scientific
    libraries. 
    The timing percentage of \texttt{walker} 
    is 62.15\% with the SVE-disabled version of DCA++ in overall program execution and
    is reduced to 40.14\% with the SVE vectorization version. 
    The percentage reduction of \texttt{walker} is similar
    to the percentage reduction of scientific libraries observed in Fig.~\ref{fig:timing_breakdown}. 
\end{enumerate}

The results show that to further improve the DCA++ application, the focus must be on tuning the application source code, particularly the accumulator code, to determine which loops need further optimization and which were successfully vectorized by the compiler. This requires significant interaction with the LLVM tools to understand the application hot spots and the opportunities for SVE optimizations. 

\section{An LLVM Tool Methodology to Generate Efficient Vectorization}

A64FX performance is highly dependent on how well the source can be mapped to
SVE instructions. It is important to determine which application loops are not being vectorized and their impact on the application's overall
performance. The ARM C/C++ compiler is based on the LLVM/Clang compiler, which
is also the basis for the authors' exploration and automation toward vectorizing the most 
important loops in an application.

Like most modern compilers, LLVM/Clang and its derivatives support profile guided optimization (PGO).
The idea is that the compiler inserts profiling instructions into the target binary to collect
information when the application is run.
During application shutdown, profiling information is stored on the disk for later use.
When the application is recompiled in the future, the collected profiling information is used to
drive heuristics (e.g., to determine a suitable unroll count for loops).
Such profiling also allows the compiler to approximate how much time was spent in a certain portion of code,
also referred to as \emph{code hotness}.
The latter makes PGO especially interesting to filter optimization remarks because it allows users to only view remarks
emitted for hot code regions.
Thus, with PGO, users can be guided toward the loops that would benefit the most from vectorization and
avoid overloading them with a plethora of uninteresting remarks.


The authors manually analyzed several loops in the DCA++ application by using the aforementioned method described to
determine what was hindering loop vectorization.
Some loops required a simple change in vectorization flags, 
and others required user intervention (e.g., vectorization directives, such as OpenMP SIMD) 
to assist the compiler.
The authors also identified loops that required transformations to make the vectorization more efficient.
The following sections present a brief discussion for four hot loops that the compiler was unable to vectorize without user intervention.

\subsection{OpenMP SIMD}

When optimizing any loops, the compiler's vectorization pass must preserve the
semantics of the original source code. 
This usually requires static analyses to verify that the transformation is legal.
However, it is not uncommon for a transformation to be correct but unable to be statically verified by the compiler.
Since OpenMP 4.0, OpenMP has added support for the SIMD directive, which provides a cross-platform method for statically
asserting information about the program's semantics to the compiler's vectorization pass~\cite{huber2017effective}.
In DCA++, various loops require additional information to be successfully vectorized.

Figure~\ref{fig:reduction} shows a classical reduction loop.
Because \lstinline|x_val| is a floating point value, any reordering of the iterations (e.g., as part of vectorization)
would break strict Institute of Electrical and Electronics Engineers (IEEE) floating point compliance and might introduce errors in the result.
By default, LLVM/Clang will not vectorize the loop but will instead emit a remark (lower part) that explains how \emph{ffast-math} or
vectorization pragmas can be used to overwrite the IEEE floating point semantics.
The Clang pragmas are a less feature-rich variant of the cross-platform OpenMP
SIMD directives, but both explicitly tell the compiler to allow vector execution for a loop.
In the OpenMP variant, users should make the parallel reduction explicit.
Additionally, the authors used the aligned clause to pass alignment information to the compiler, which can lead to improved
performance due to specialized memory instructions.

\begin{figure}[ht]
\vspace*{-2mm}
\centering
\begin{lstlisting}
#pragma omp simd reduction(-:x_val) aligned(x_val, G_ptr : 64)
for (int i = 0; i < j; i++)
  x_val -= x_ptr[i] * G_ptr[i];
\end{lstlisting}
\vspace*{-2mm}
\begin{lstlisting}[numbers=none,breakindent=0pt,language=]
remark: loop not vectorized: cannot prove it is safe to reorder floating-point operations; allow reordering by specifying '#pragma clang loop vectorize(enable)' before the loop or by providing the compiler option '-ffast-math'
\end{lstlisting}
\vspace*{-5mm}
\caption{A loop performing a parallel reduction that is not vectorized automatically.}
\label{fig:reduction}
\vspace*{-2mm}
\end{figure}

In line~\ref{line:gather} of Figure~\ref{fig:gather}, there is a noncontinuous memory load---a gather.
ARM's SVE supports fast gathering operations; however, the compiler cannot vectorize this loop without manual intervention
because the accessed arrays \lstinline|M_ij_|, \lstinline|M|, \lstinline|config_left_|, and \lstinline|config_right_| might alias and hence overlap.
In these situations, the compiler is often able to version the loop and generate a vectorized variant guarded by a run time alias check to verify that the accessed ranges of the arrays do not overlap at run time.
However, the support for such run time alias checks in LLVM/Clang is limited to the case in which the accessed bounds are known statically~\cite{runtime_alias_check}.
Because the index into the \lstinline|M| array is based on the values loaded from the configuration arrays, the access range cannot be bound statically.
The compiler remark shown below the loop nest summarizes this discussion in a way that is difficult or impossible for application developers to understand.
Using OpenMP SIMD effectively tells the compiler that there are no overlapping accesses, allowing the loop to be vectorized.
Care must be taken to ensure that no aliasing actually occurs, otherwise this will result in incorrect results.

\begin{figure}[ht]
\centering
\begin{lstlisting}
for (int j = start_index_right_[orb_j]; j < end_index_right_[orb_j]; ++j) {
  const int out_j = j - start_index_right_[orb_j];
  #pragma omp simd
  for (int i = start_index_left_[orb_i]; i < end_index_left_[orb_i]; ++i) {
    const int out_i = i - start_index_left_[orb_i];
    M_ij_(out_i, out_j) = M(config_left_[i].idx, config_right_[j].idx); <@\label{line:gather}@>
  }
}
\end{lstlisting}
\vspace*{-2mm}
\begin{lstlisting}[numbers=none,breakindent=0pt,language=]
remark: loop not vectorized: Unknown array bounds
\end{lstlisting}
\vspace*{-5mm}
\caption{A loop performing a memory gather that requires OpenMP SIMD to be
  vectorized by the ARM compiler.}
\label{fig:gather}
\end{figure}

\subsection{Using the Correct Compiler Flags}

Some loops require additional compiler flags to be vectorized. 
The code shown in Figure~\ref{fig:library} has two run time calls, line~\ref{line:math} and \ref{line:math2}, which prevent the compiler
from automatically vectorizing it.
A function call usually requires an explicit vector version of the function and compiler support to allow vectorized execution.
The ARM compiler provides an optimized math library that includes vector variants of common math functions.
Users must explicitly enable such a vector library because it will disturb the precision of the result, similar to the floating point reordering.
The ARM compiler provides the \emph{fsimdmath} option to use its performance libraries, whereas standard Clang requires \emph{fveclib} to be set to the desired vectorized library.
\emph{ffast-math} or \emph{fno-math-errno} will allow the compiler to execute the loop out of order, but no vectorized math library is used.
This means that the vector lanes are effectively unpacked before the call, and the math function is executed once per vector lane.

Another issue is that the application uses a custom matrix class that performs bounds checking by using assertions in the overloaded access operators.
Although assertions are a good software engineering practice, their ``complex'' semantics must be preserved by the compiler.
The problem is that no code is executed after a violated assertion.
Thus, if assertions are enabled and present in a loop, the compiler must verify that the assertion cannot trigger to execute
any side effects succeeding the assertion (e.g., from the next iteration).
To disable assertions completely, \emph{NDEBUG} can be defined during compilation; however this will cause a tension between ``debug'' and ``release'' builds that is often not desirable.
For developers to identify issues that stem from assertions and other errors in handling code, the authors added a new remark to the LLVM vectorizer, which is shown below the code.
For these experiments, the authors disabled assertions, provided a vectorized math library, and added OpenMP SIMD to allow vectorization, even in the presence of possibly aliasing accesses. 

\begin{figure}[ht]
\centering
\begin{lstlisting}
for (int j = 0; j < n_v; ++j) {
#pragma omp simd
  for (int i = 0; i < n_w; ++i) {
    const ScalarType x = configuration[j].get_tau() * w_[i];
    T_[0](i, j) = std::cos(x); <@\label{line:math}@>
    T_[1](i, j) = std::sin(x); <@\label{line:math2}@>
  }
}
\end{lstlisting}
\vspace*{-2mm}
\begin{lstlisting}[numbers=none,breakindent=0pt,language=]
remark: loop not vectorized: loop exit block contains control flow that does not return
remark: loop not vectorized: library call cannot be vectorized. Try compiling with -fno-math-errno, -ffast-math, or similar flags
\end{lstlisting}
\vspace*{-5mm}
\caption{A code block using the math library functions \emph{cos} and
  \emph{sin}.}
\label{fig:library}
\end{figure}

\subsection{Loop Transformations}

The loop in Fig.~\ref{fig:transform} contains gathers
from memory at lines~\ref{line:gather1}~and~\ref{line:gather2}. More importantly, the code uses a column-major layout for all its matrices while this loop iterates across a row. This will require
expensive scattering operations to distribute the stores to discontinuous
memory addresses. This loop can be transformed to better exploit SIMD parallelism. Each iteration of this loop is independent, and the matrices
are guaranteed to be square in the code, so this loop can safely be transposed to improve memory accesses. This transformation will also improve performance without vectorizing the loop.

\begin{figure}[ht]
\centering
\begin{minipage}[t]{0.50\columnwidth}
\begin{lstlisting}[basicstyle=\scriptsize]
for (int i = 0; i < Gamma.Rows(); i++) {
  for (int j = 0; j < Gamma.Cols(); j++) {
    int spin_idx_i = random_vertex_vector[i];
    int spin_idx_j = random_vertex_vector[j];

    if (spin_idx_j < vertex_index) {<@\label{line:branch}@>
      Real delta = (spin_idx_i == spin_idx_j)
                       ? 1.
                       : 0.;
      Real N_ij = N(spin_idx_i, spin_idx_j);
      Gamma(i, j) =
          (N_ij * exp_V[j] - delta) /
          (exp_V[j] - 1.);
    } else
      Gamma(i, j) = G_precomputed(
          spin_idx_i,
          spin_idx_j - vertex_index);<@\label{line:cond}@>
    if (i == j) { 
      Real gamma_k = exp_delta_V[j];
      Gamma(i, j) -=
          (gamma_k) / (gamma_k - 1.);
    }
  }
}
\end{lstlisting}

\end{minipage}
\hskip -12pt
\hfill
\begin{minipage}[t]{0.50\columnwidth}
\begin{lstlisting}[basicstyle=\scriptsize,numbers=none]
for (int j = 0; j < Gamma.Cols(); j++) {
#pragma omp simd
  for (int i = 0; i < Gamma.Rows(); i++) {
    int spin_idx_i = random_vertex_vector[i];
    int spin_idx_j = random_vertex_vector[j];

    if (spin_idx_j < vertex_index) {
      Real delta = (spin_idx_i == spin_idx_j)
                       ? 1.
                       : 0.;
      Real N_ij = N(spin_idx_i, spin_idx_j); <@\label{line:gather1}@>
      Gamma(i, j) =
          (N_ij * exp_V[j] - delta) /
          (exp_V[j] - 1.); <@\label{line:div}@>
    } else
      Gamma(i, j) = G_precomputed(
          spin_idx_i,
          spin_idx_j - vertex_index); <@\label{line:gather2}@>
  }
  
  Real gamma_k = exp_delta_V[j];
  Gamma(j, j) -= 
      (gamma_k) / (gamma_k - 1.);
}
 
\end{lstlisting}
\end{minipage}
\vspace*{-2mm}
\begin{lstlisting}[numbers=none,breakindent=0pt,language=]
remark: loop not vectorized: control flow cannot be substituted for a select
remark: loop not vectorized: cannot identify array bounds 
\end{lstlisting}
\vspace*{-5mm}
\caption{A loop requiring a source transformation and OpenMP SIMD (left) and its transformed version (right).}
\label{fig:transform}
\end{figure}

This loop contains conditional expressions that must be transformed into masks
to be vectorized. This requires calculating the result of each branch and
conditionally moving it into the final register by using a mask. In this case, the true condition of the loop at line~\ref{line:branch} is much more
computationally expensive than the false condition. If the result was not
needed, then this will be calculated at each iteration of the loop, only to be thrown
away. This problem is even worse for the final update across the diagonal at line~\ref{line:cond}, which will only be needed once every iteration of the inner loop but
calculated every iteration. This conditional update can be hoisted from the loop to improve performance significantly.

Another issue is the division at line~\ref{line:div}. This could cause a division-by-zero error that can block
vectorization if regular error handling semantics are maintained. This can be
disabled with fast math, but in some cases, the compiler is able to vectorize it by using masked division instructions. This would be a good application of the assume
directive added in OpenMP 5.1 to assert to the compiler that the division will never cause an error.

\subsection{Results}

\begin{figure}[ht]
    \centering
    \includegraphics[width=0.8\columnwidth, trim=0.1cm 0.1cm 0.1cm 0.1cm, clip]{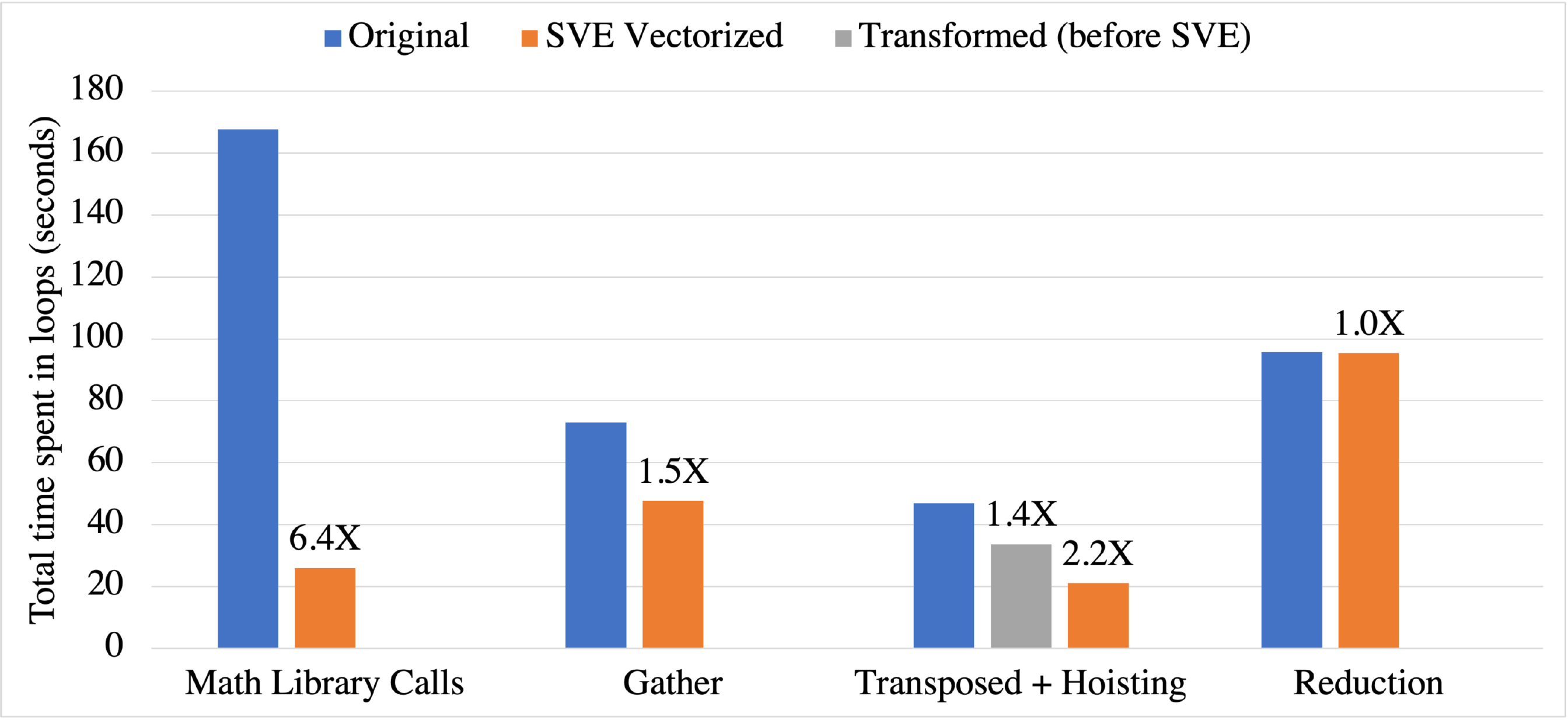}
    \caption{The loops in Figs.~\ref{fig:library},~\ref{fig:gather},~\ref{fig:transform},~and \ref{fig:reduction}, respectively, before and after the barriers to SVE execution were remedied. Performance is measured as the total time spent by all the threads in a run using 24 accumulators/walker threads over $100,000$ measurements.}
    \label{fig:looptiming}
\end{figure}

The overall impact of these transformations is shown in
Fig.~\ref{fig:looptiming}, which shows a significant speedup in most cases. The loop in Fig.~\ref{fig:library} had the largest improvement when using ARM's vector math support. The
reduction loop in Fig.~\ref{fig:reduction} yielded no improvement. Upon
further investigation, this was because the loop's trip count was very small in
the average case, so the majority of the time was spent doing the final reduction, and work was rarely done in parallel. The other loops saw reasonable improvements, but their performance was limited by the gathering instructions required to vectorize them.

\section{Automating the Process: The OpenMP Advisor}

It is unrealistic but unfortunately still common practice to optimize code and add support for new platforms and features by manually inspecting and modifying the application.
Given the increasing complexity when it comes to hardware and the requirement to support multiple heterogeneous platforms simultaneously, the authors must rethink their software engineering practices to ensure that the code is not only correct but also performant and portable.
To automate this manual process and boost programmers' productivity, the authors began developing the OpenMP Advisor.
Based on the portable OpenMP directive language, we hope to evolve the OpenMP Advisor over time into a valuable software engineering tool by using and extending LLVM capabilities.
During the porting effort of the DCA++ application described here, the authors experienced various issues that require interpretation to derive actionable advice.
Using their experience, the authors began automating the parts of the process and improving the compiler remarks that were missing or misleading.
As a result, the OpenMP Advisor the authors develop as part of the LLVM compiler framework will use optimization remarks from multiple optimization passes to report the most performance-critical problems in the code based on the available profiling data.

\section{Related Work}
There are several other tools that analyze source code or provide support for parallelization but with limited support that automatically inserts SIMD directives in the code. These include: CAPO~\cite{IEROTHEOU2005999} for automatic OpenMP work-sharing directives generation, which supports Fortran 77 and some F90 extensions; Appentra's Parallware~\cite{10.1007/978-3-030-34356-9_27}, which focuses on parallelizing C/C++ applications by using OpenMP and OpenACC for multicores and accelerators; and Cray Reveal~\cite{DEROSE20141480}, which helps autoscope OpenMP variables and generate OpenMP work-sharing for Fortran and C/C++ for multicore and accelerators. Intel Inspector focuses on OpenMP semantic checking for data race detection. Foresys~\cite{Pazat1996} and the Dragon Analysis tool~\cite{1236416} are legacy tools that supported the maintenance of Fortran code and help with parallelization with OpenMP. 

\section{Conclusion}
Porting the DCA++ application to the A64FX processor requires the use of optimized scientific libraries and vectorizing the application hot spots. This process can be overwhelming to users, and tools are needed to automate this process. This work shows that by using LLVM tools, users can easily detect hot spots, determine why loops are not vectorized, and correct the issues by applying the correct compiler flags, transforming the code, or applying OpenMP directives.

Currently, authors are working an OpenMP Advisor tool that is built on top of existing and newly introduced LLVM tooling to automate this process. 
Ultimately, the authors want to enable application developers to navigate and handle compiler-generated information productively.
Optimization reports should pinpoint important opportunities to tune the code (e.g., non-vectorized loops) and simultaneously provide sufficient information and suggestions to allow informed decisions without elaborate studies of compiler and programming language theory.
The authors believe that tools can recommend portable annotations, such as OpenMP SIMD directives, when they inform users about the requirements for correctness.
Furthermore, compiler analysis and optimizations can directly target the recently proposed OpenMP assume directive to request user feedback.
In other words, OpenMP assume directives and the authors' implementation in the LLVM compiler will enable analyses and transformations to request high-level information from users naturally.
The OpenMP Advisor will improve communication in the other direction to present users with important requests and remarks, together with information and examples that translate ``compiler language'' to ``application language.''

\section*{Acknowledgment}
The authors would like to thank Manuel Arenaz (Appentra Solutions), Hartmut Kaiser (Louisiana State University), and Kevin Huck (University of Oregon) for their guidance and 
feedback on this work.

This work was supported by the Scientific Discovery through Advanced Computing (SciDAC) 
program funded by US Department of Energy, Office of Science, Advanced Scientific Computing 
Research (ASCR) and Basic Energy Sciences (BES) Division of Materials Sciences and 
Engineering. This research was also supported by the Exascale Computing Project (17-SC-20-SC), a collaborative effort of the US Department of Energy Office of Science and the National Nuclear Security Administration, in particular its subproject on Scaling OpenMP with LLVM for Exascale performance and portability (SOLLVE). 

Notice: This manuscript has been authored by UT-Battelle, LLC, under contract DE-AC05-00OR22725 with the US Department of Energy (DOE). The US government retains and the publisher, by accepting the article for publication, acknowledges that the US government retains a nonexclusive, paid-up, irrevocable, worldwide license to publish or reproduce the published form of this manuscript, or allow others to do so, for US government purposes. DOE will provide public access to these results of federally sponsored research in accordance with the DOE Public Access Plan (\url{http://energy.gov/downloads/doe-public-access-plan}).

This work was performed under the auspices of the U.S. Department of Energy by Lawrence Livermore National Laboratory under Contract DE-AC52-07NA27344 (LLNL-CONF-819815).

\bibliography{references}

\begin{thebibliography}{10}

\bibitem{hahner2020dca++}
Urs~R H{\"a}hner, Gonzalo Alvarez, Thomas~A Maier, Raffaele Solc{\`a}, Peter
  Staar, Michael~S Summers, and Thomas~C Schulthess.
\newblock Dca++: A software framework to solve correlated electron problems
  with modern quantum cluster methods.
\newblock {\em Computer Physics Communications}, 246:106709, 2020.

\bibitem{dca_hpx_2020}
Weile Wei, Arghya Chatterjee, Kevin Huck, Oscar Hernandez, and Hartmut Kaiser.
\newblock Performance analysis of a quantum monte carlo application on multiple
  hardware architectures using the hpx runtime.
\newblock In {\em 2020 IEEE/ACM 11th Workshop on Latest Advances in Scalable
  Algorithms for Large-Scale Systems (ScalA)}, pages 77--84. IEEE, 2020.

\bibitem{Kaiser2020}
Hartmut Kaiser, Patrick Diehl, Adrian~S. Lemoine, Bryce~Adelstein Lelbach,
  Parsa Amini, Agustín Berge, John Biddiscombe, Steven~R. Brandt, Nikunj
  Gupta, Thomas Heller, Kevin Huck, Zahra Khatami, Alireza Kheirkhahan, Auriane
  Reverdell, Shahrzad Shirzad, Mikael Simberg, Bibek Wagle, Weile Wei, and
  Tianyi Zhang.
\newblock Hpx - the c++ standard library for parallelism and concurrency.
\newblock {\em Journal of Open Source Software}, 5(53):2352, 2020.

\bibitem{huck2015autonomic}
Kevin~A Huck, Allan Porterfield, Nick Chaimov, Hartmut Kaiser, Allen~D Malony,
  Thomas Sterling, and Rob Fowler.
\newblock An autonomic performance environment for exascale.
\newblock {\em Supercomputing frontiers and innovations}, 2(3):49--66, 2015.

\bibitem{papi2010}
Dan Terpstra, Heike Jagode, Haihang You, and Jack Dongarra.
\newblock Collecting performance data with papi-c.
\newblock In {\em Tools for High Performance Computing 2009}, pages 157--173.
  Springer, 2010.

\bibitem{diehl2021performance}
Patrick Diehl, Dominic Marcello, Parsa Armini, Hartmut Kaiser, Sagiv Shiber,
  Geoffrey~C. Clayton, Juhan Frank, Gregor Daiß, Dirk Pflüger, David Eder,
  Alice Koniges, and Kevin Huck.
\newblock Performance measurements within asynchronous task-based runtime
  systems: A double white dwarf merger as an application, 2021.

\bibitem{huber2017effective}
Joseph~N. Huber, Oscar~R. Hernandez, and Matthew~Graham Lopez.
\newblock Effective vectorization with openmp 4.5.
\newblock 3 2017.

\bibitem{runtime_alias_check}
P\'{e}ricles Alves, Fabian Gruber, Johannes Doerfert, Alexandros Lamprineas,
  Tobias Grosser, Fabrice Rastello, and Fernando Magno Quint\~{a}o Pereira.
\newblock Runtime pointer disambiguation.
\newblock In {\em Proceedings of the 2015 ACM SIGPLAN International Conference
  on Object-Oriented Programming, Systems, Languages, and Applications}, OOPSLA
  2015, page 589–606, New York, NY, USA, 2015. Association for Computing
  Machinery.

\bibitem{IEROTHEOU2005999}
C.S. Ierotheou, H.~Jin, G.~Matthews, S.P. Johnson, and R.~Hood.
\newblock Generating openmp code using an interactive parallelization
  environment.
\newblock {\em Parallel Computing}, 31(10):999--1012, 2005.
\newblock OpenMP.

\bibitem{10.1007/978-3-030-34356-9_27}
Manuel Arenaz and Xavier Martorell.
\newblock Parallelware tools: An experimental evaluation on power systems.
\newblock In Mich{\`e}le Weiland, Guido Juckeland, Sadaf Alam, and Heike
  Jagode, editors, {\em High Performance Computing}, pages 352--360, Cham,
  2019. Springer International Publishing.

\bibitem{DEROSE20141480}
Luiz DeRose, Heidi Poxon, James Beyer, and Alistair Hart.
\newblock A high level programming environment for accelerator-based systems.
\newblock {\em Procedia Computer Science}, 29:1480--1490, 2014.
\newblock 2014 International Conference on Computational Science.

\bibitem{Pazat1996}
Jean-Louis Pazat.
\newblock {\em Tools for high performance fortran: A survey}, pages 134--158.
\newblock Springer Berlin Heidelberg, Berlin, Heidelberg, 1996.

\bibitem{1236416}
B.~{Chapman}, O.~{Hernandez}, {Lei Huang}, {Tien-hsiung Weng}, {Zhenying Liu},
  L.~{Adhianto}, and {Yi Wen}.
\newblock Dragon: an open64-based interactive program analysis tool for large
  applications.
\newblock In {\em Proceedings of the Fourth International Conference on
  Parallel and Distributed Computing, Applications and Technologies}, pages
  792--796, 2003.

\end{thebibliography}
\bibliographystyle{unsrt}  

\end{document}